# A simple modeling for gas release during annealing of irradiated nuclear fuel

J. Losfeld*, L. Desgranges*, Y. Pontillon*, and G. Baldinozzi**

*DEC/SA3E/LAMIR, CEA, DES, IRESNE, Cadarache,
13108 Saint-Paul-lez-Durance, France (e-mail: jimmy.losfeld@cea.fr).
**Laboratoire Structures, Propriétés et Modélisation des Solides,
CNRS-UMR 8580 Université Paris-Saclay, Centralesupélec, CNRS, SPMS, 91190 Gif-sur-Yvette, France

## INTRODUCTION

In irradiated $UO_2$, fission gases accumulate in the fuel and form different reservoirs. The size and location of these reservoirs depend on many parameters, such as burn-up and power history. During an annealing experiment of an irradiated fuel sample, these reservoirs change, interconnect, and under the effect of the thermal load can give rise to a release outside the sample. This corresponding fission gas release (FGR) occurs in successive bursts at different temperature levels. Although the mechanisms to explain this release are not completely identified, the literature suggests that the grain boundaries decohesion is the preferred mechanism. [1]

More specifically, annealing test results on irradiated $UO_2$ sample from work by Noirot et al. [2] and Marcet et al. [3] indicates that the release around 1100°C comes preferentially from the HBS zone of the fuel. This temperature level coincides with the last burst of release (Burst around 1200°C in figure 2) highlighted in slow ramp annealing performed in the MERARG furnace on high burn up irradiated fuel samples. [4] We aim to determine whether the release kinetics observed at these temperature levels are consistent with a gas reservoir located in the intergranular bubbles of the HBS zone. In this paper, we propose a simplified model of the gas flows of the third burst in order to determine the physical characteristics of the exit channels followed by the fission gas.

## Modelling fission gas emptying kinetics

The release of the gas contained in the HBS bubbles takes place through their interconnection to a free surface, as can be seen in the SEM image in Figure 1.

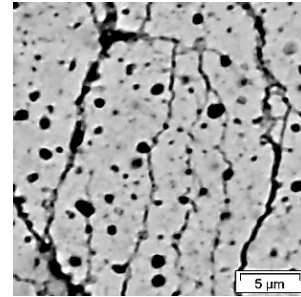

Fig. 1. Post annealing SEM image of the HBS area of an irradiated $UO_2$ fuel sample

We have developed a gas flow model in the fuel during the annealing. It postulates that the gas release during an isothermal plateau at 1200°C corresponds to the equilibrium between overpressure gases reservoirs located in the fuel sample connected to the free surface at atmospheric pressure. We will simplify the gas reservoir corresponding to the intergranular gas bubbles interconnection of the HBS zone as a single reservoir connected by a single channel outside of the fuel as shown in Figure 2.

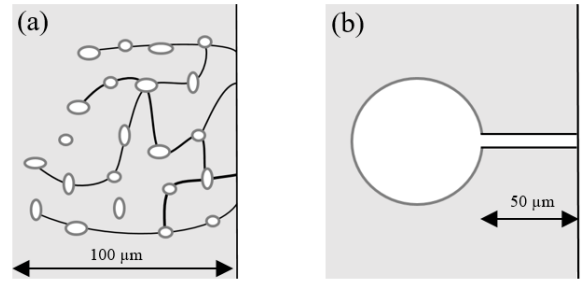

Fig. 2. Schematic of the interconnection between the sample's actual (a) and simulated (b) fission gas reservoirs to the free surface.

Thus defined, and from the mass balance, Darcy and perfect gas equations, the release kinetics of krypton 85 during the isothermal stage should take the form of equation (1).

FGR kinetics equation during discharge:

$$FGR_{Kr85} = \lambda \times \left(\left(\frac{2 \cdot P_e}{C\, e^{-\frac{t}{\tau}} - 1} + P_e\right)^2 - P_e^2\right) \quad (1)$$

With:
$\lambda$ : scale factor
$P_e$: external pressure
$\tau$ : decay factor (inversely proportional to the channel permeability)
C : factor related to the initial pressure of the gas reservoir

The annexes contain the details of the calculation to obtain this equation.

**Experimental data:**

We tested our model on an annealing test on a highly irradiated $UO_2$ sample with a temperature ramp of 0.2°C/s from 300 to 1200°C as shown in Figure 3. At 1200°C we maintained an isothermal plateau until the end of the gas release was observed. During the ramp the release describes different bursts that we can interpret as resulting from the successive opening of the release channels due to the increase of the pressure of the associated gas reservoirs. From the 100th minute, we observe the third burst, which we hypothesised to be correlated with the HBS zone of the sample. From the beginning of the 1200°C isothermal plateau, we observe the change of the gas release regime, and its decrease. We will focus on the characteristic decay of the gas release curve from the 125th minute, as highlighted in the insert to Figure 3, when the gas release regime is established, which we will fit with the above-mentioned model.

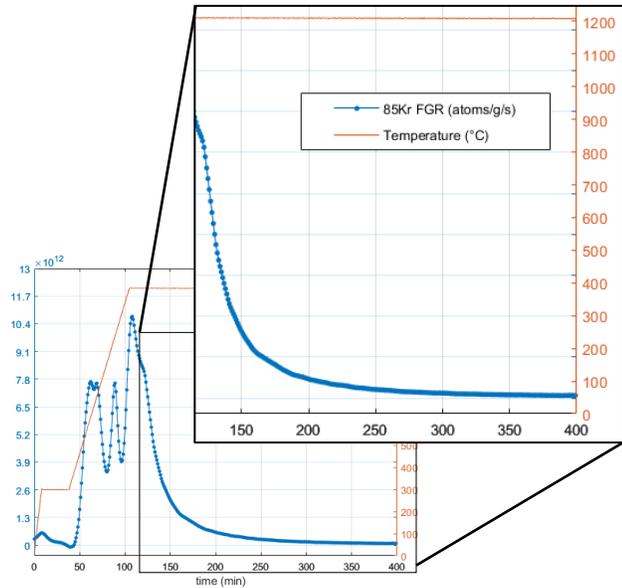

Fig. 3. Example of emptying kinetics observed at the 125th minute on the release curve of krypton 85 during annealing.

**Dataset fitting**

From equation (1), the fit involves the following three parameters: $\lambda$, $\tau$ and **c**
These three parameters are highly dependent, which gives fitting results that are very sensitive to small variations. To overcome this sensitivity and improve the robustness of the fit, we will use the logarithmic derivative method.

The main advantage of using the logarithmic derivative in our fit is to simplify the number of parameters to be fit by simplifying the scale factor $\lambda$, which does not interest us here. Thus, the fit will concern the parameters: $\tau$ and **c**

The second benefit of the logarithmic derivative is that it allows us to identify the beginning of the emptying phase, i.e. the inflection point of the emptying curve where we will place $t_0$.

Therefore:

$$Dlog\,(RGF_{Kr85}) = \frac{4 \cdot C \cdot P_e \cdot \exp\left(-\frac{t}{\tau}\right)}{\tau \cdot \left(C \cdot \exp\left(-\frac{t}{\tau}\right) - 1\right)^2} \times \left(\frac{2 \cdot P_e}{C \cdot \exp\left(-\frac{t}{\tau}\right) - 1} + P_e\right) \quad (2)$$

We have applied the simplified gas flow model defined above to the example of emptying kinetics shown in Figure 3. It fits the experimental data accurately as shown in figure 4. The model is well suited to annealing where an isothermal stage at 1200°C is maintained long enough to observe the emptying kinetics. The question now arises as to how to interpret the model. As we do not know how many paths are created during discharge, we reason here on the scale of an equivalent channel, i.e. if the emptying had only taken place in a single path. Our calculations allow us to determine the equivalent permeability (K) of the channel from $\tau$ as follows:

$$K = \sqrt{-\frac{\mu \cdot L \cdot V_i}{P_e \cdot 8\,\pi\,\tau}} \quad (3)$$

$\mu$, $P_e$ et $\tau$ are known. $V_i$ the volume of the reservoir associated with the emptying third burst can be measured experimentally. We know that the gas released by this burst is mainly from the HBS zone, so we will assume that the length of the emptying path will be of the same order of magnitude as the width of the HBS of the sample, i.e. 50 µm.

Thus, for the example studied, we determined the equivalent permeability as being equal to $1{,}28 \cdot 10^{-13}$ m². Given our hypotheses, we can interpret this equivalent permeability value as a maximum value. Indeed, the higher the number of emptying channels, the lower the value of the permeability of these channels will be.

To conclude, without going further in terms of interpretation, it is interesting to note that the order of magnitude of equivalent permeability obtained is close to the permeability of porous rocks such as sandstone or granite, studied in geology.[5] This encouraging outcome encourages us to improve the model in order to refine the results.

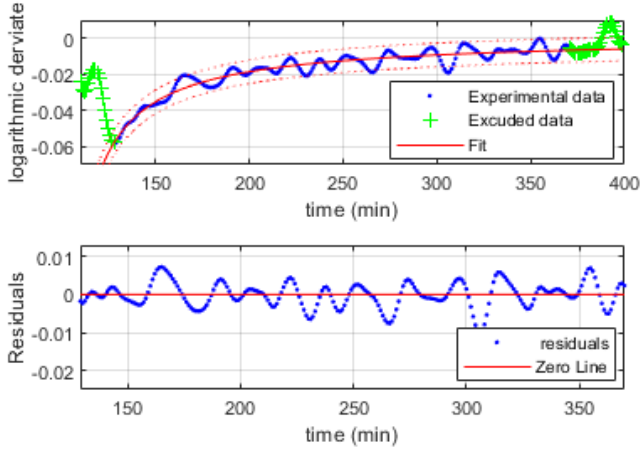

Fig. 4. Logarithmic derivative fitting (top) and residuals (bottom) applied to the release kinetics shown in Figure 1.

## ANNEXES

### Model of the fission gas emptying into the fuel

This description consists of a mass balance, Darcy's law and the law of perfect gases:

- Mass balance

$$\frac{\partial}{\partial t}(\omega\,\rho) + \text{div}(\rho u) = 0 \quad (4)$$

The mass balance expresses the variation of the gas density ρ [kg.m-3], as a function of its velocity u [m.s-1]. It depends on a first parameter of the porous medium: its porosity ω [without unit].

- Darcy

$$G = u.S = -\frac{\overline{K}}{\mu}.S.\,\text{grad}(p) \quad (5)$$

Darcy's law expresses the flow of a gas G [m3/s], product of the velocity u [m/s] of gases and an area S [m²], as a function of pressure gradients p [Pa]. It depends on the permeability tensor K [m²], which characterises the porous medium from the point of view of flow, and the dynamic viscosity μ [kg.s-1 .m-1] of the gas.

- Perfect gas

$$\rho = p\,\frac{M}{R\,T} \quad (6)$$

Expression of the density, as a function of the perfect gas constant R=8.31 J.mol-1 .K-1, M [kg/mol] the molar mass of the gas, and T [K] the temperature.

Injecting (5) and (6) into (4) we obtain:

$$\frac{\partial}{\partial t}\left(\omega\,\frac{p}{T}\right) = \text{div}\left(\frac{p}{T}.\frac{\overline{K}}{\mu}.\,\text{grad}(p)\right) \quad (7)$$

### Isothermal flow:

For isothermal flow, the constant temperature can be simplified on either side of the above expression. Applied to our example it will take this form:

$$V_i\,\frac{\partial P_i}{\partial t} = -\frac{K.S}{2.\mu.L} \times \left(P_i^2 - P_e^2\right) \quad (8)$$

With :
$V_i$ : the volume of gas inside the sample
$P_i$: the pressure inside the gas tank
$P_e$: the pressure in the oven, here constant and equal to atmospheric pressure
$K$: the permeability of the channel
S: the cross-sectional area of the channel
L: the length of the channel
µ: the dynamic viscosity of the gas

By integrating equation 8 we obtain the time expression of $P_i$ :

$$P_i(t) = \frac{2 \cdot P_e}{C e^{-\frac{t}{\tau}} - 1} + P_e \quad (9)$$

With $C = 1 + \frac{2}{\frac{P_0}{P_e} - 1}$ and $\tau = -\frac{\mu \cdot L \cdot V_i}{P_e \cdot K \cdot S}$

From the time expression of Pi, we can determine the expression of the gas outlet flow with Darcy's law:

$$Q = \frac{K \cdot S}{2 \cdot \mu \cdot L} \times (P_i^2 - P_e^2) \quad (10a)$$

$$Q = \frac{K \cdot S}{2 \cdot \mu \cdot L} \times \left( \left( \frac{2 \cdot P_e}{C e^{-\frac{t}{\tau}} - 1} + P_e \right)^2 - P_e^2 \right) \quad (10b)$$

In MERARG we measure a quantity of 85Kr atoms relative to the mass of the sample, not a total flux. To simplify and adapt equation 10 to our release data, we will use the following equation to model the release of 85Kr fission gas:

$$RGF_{Kr85} = \lambda \times \left( \left( \frac{2 \cdot P_e}{C e^{-\frac{t}{\tau}} - 1} + P_e \right)^2 - P_e^2 \right) \quad (1)$$